\documentclass[aps,prl,superscriptaddress,twocolumn]{revtex4}

\usepackage[pdftex]{graphicx}
\usepackage{dcolumn}
\usepackage{bm}
\usepackage{color}
\usepackage{amsmath}
\usepackage{nicefrac}
\usepackage{amssymb} 

\newcommand{\ff}[1]{{\boldsymbol #1}}

\newcommand{\bi}{\begin{itemize}}
\newcommand{\ei}{\end{itemize}}
\newcommand{\be}{\begin{equation}}
\newcommand{\ee}{\end{equation}}
\newcommand{\ba}{\begin{eqnarray}}
\newcommand{\ea}{\end{eqnarray}}
\newcommand{\nab}{\boldsymbol \nabla}

\usepackage[pdftex,colorlinks=true,linkcolor=blue,citecolor=blue,filecolor=blue]{hyperref}

\begin{document} 
  
\title{Topological spin torque emerging in classical-spin systems with different time scales}

\author{Michael Elbracht} 

\affiliation{I. Institute of Theoretical Physics, Department of Physics, University of Hamburg, Jungiusstra\ss{}e 9, 20355 Hamburg, Germany}

\author{Simon Michel}

\affiliation{I. Institute of Theoretical Physics, Department of Physics, University of Hamburg, Jungiusstra\ss{}e 9, 20355 Hamburg, Germany}

\author{Michael Potthoff}

\affiliation{I. Institute of Theoretical Physics, Department of Physics, University of Hamburg, Jungiusstra\ss{}e 9, 20355 Hamburg, Germany}

\affiliation{The Hamburg Centre for Ultrafast Imaging, Luruper Chaussee 149, 22761 Hamburg, Germany}

\begin{abstract}
In classical spin systems with two largely different inherent time scales, the configuration of the fast spins almost instantaneously follows the slow-spin dynamics. 
We develop the emergent effective theory for the slow-spin degrees of freedom and demonstrate that this generally includes a topological spin torque. 
This torque gives rise to anomalous real-time dynamics. 
It derives from the holonomic constraints defining the fast-spin configuration space and is given in terms of a topological charge density which becomes a quantized homotopy invariant when integrated.
\end{abstract} 

\maketitle 

\section{Introduction}
\label{sec:intro}

\paragraph{\color{blue} Introduction.}  

Topological charges are homotopy invariants \cite{Nak98}, which can take discrete values only and which are used in theoretical physics to discriminate between topologically different states of physical systems. 
They have supplemented the more established concept of classifying states of matter based on symmetry, spontaneous symmetry breaking and order parameters.
Well known topological invariants for gapped quantum systems are Chern numbers or Z$_{2}$ invariants classifying, e.g., topological insulators \cite{HK10,QZ11}.
For classical systems, e.g., for anisotropic classical-spin systems, skyrmion numbers \cite{Bra12,NT13,FBT+16} are used to characterize topologically different magnetic states.

Quantum-mechanically as well as classically, these topological charges derive from {\em locally} defined and gauge invariant quantities, i.e., from topological charge {\em densities} describing local properties of vector bundles. 
A prime example is given by the Berry curvature, which is a phase 2-form in parameter space \cite{Sim83}.
In particular, these topological charge densities describe the effect on the system's state when steering the system along a closed path in parameter space. 
A quantum state picks up a geometrical phase, the Berry phase \cite{Ber84,BMK+03}, which is obtained by integrating the topological charge density, the Berry curvature, over a surface enclosed by the path. 
An analog for classical systems is Hannay's angle variable holonomy \cite{Han85}, which arises for integrable systems when the Hamiltonian is adiabatically taken around a closed path in parameter space.

While the effects of slowly varying parameters
on the local and the global topological properties of the system's state have been studied extensively, for gapped quantum condensed-matter systems \cite{vK86,TKNN82,HK10,QZ11},
for discrete quantum systems with degenerate eigenstates \cite{WZ84},
for quantum-spin \cite{Hal83a,Hal83b,AKLT87,PT19} and classical-spin models \cite{Bra12,NT13,FBT+16,VA14},
in the context of classical phase transitions \cite{Kos74,KT72} and
in molecular dynamics \cite{MH00,BMK+03}, etc.,
the {\em feedback} of the local topological charge densities {\em on the state of the parameters} has not so much been in the focus.
This feedback is meaningful, if the ``parameters'' are actually treated as classical dynamical degrees of freedom with a real-time dynamics that is slow compared to the fast degrees of freedom of the ``system''.

\paragraph{\color{blue} Anomalous slow-spin dynamics.}

With the present Letter we would like to adopt this change of the perspective. 
We consider the real-time dynamics of a purely classical system which is governed by two largely different intrinsic time scales.
The role of the parameters is played by ``slow'' spins.
Their states define a base manifold.
Assuming that the ``fast''-spin subsystem follows the slow spins {\em adiabatically} when the slow-spin state evolves in time, we can define a topological charge density which is reminiscent of a skyrmion density \cite{Bra12,NT13,FBT+16} but for skyrmions living on the base manifold, which is given by a Cartesian product of Bloch spheres rather than by Euclidean space. 
The corresponding topological charge is quantized. 

More importantly, however, as we can demonstrate very generally, there is an effective theory involving the slow-spin degrees of freedom only, and here the topological charge density gives rise to an unconventional topological spin torque. 
This torque can lead to sizeable anomalous effects as is explicitly demonstrated for a simple toy model with a single slow classical impurity spin coupled to a classical Heisenberg model.
We also check the effective theory against the numerical solution of the full set of dynamical canonical equations. 
This pinpoints the model parameter range where the dynamics is adiabatic, i.e., Born-Oppenheimer-like \cite{MH00}.

There are a few earlier studies of the dynamical role of the Berry curvature all addressing, however, quantum-classical hybrid systems \cite{KI85,ZW06}, particularly semiclassical electron dynamics in crystals \cite{Res00} and adiabatic long-wavelength magnon dynamics \cite{WZ88,NK98,NWK+99}.
Let us emphasize that the present work discusses a conceptually much simpler class of systems, namely systems with entirely classical spin degrees of freedom as given, e.g., by standard classical Heisenberg-type models. 
Such models are ubiquituously employed, e.g., in the field of classical atomistic spin dynamics
\cite{SHNE08,TKS08,BMS09,EFC+14} modelling a great variety of magnetic phenomena, where separation of time scales is frequently caused by weakly coupled spins or by strong magnetic anisotropies. 

\begin{figure}[t]
\includegraphics[width=0.65\columnwidth]{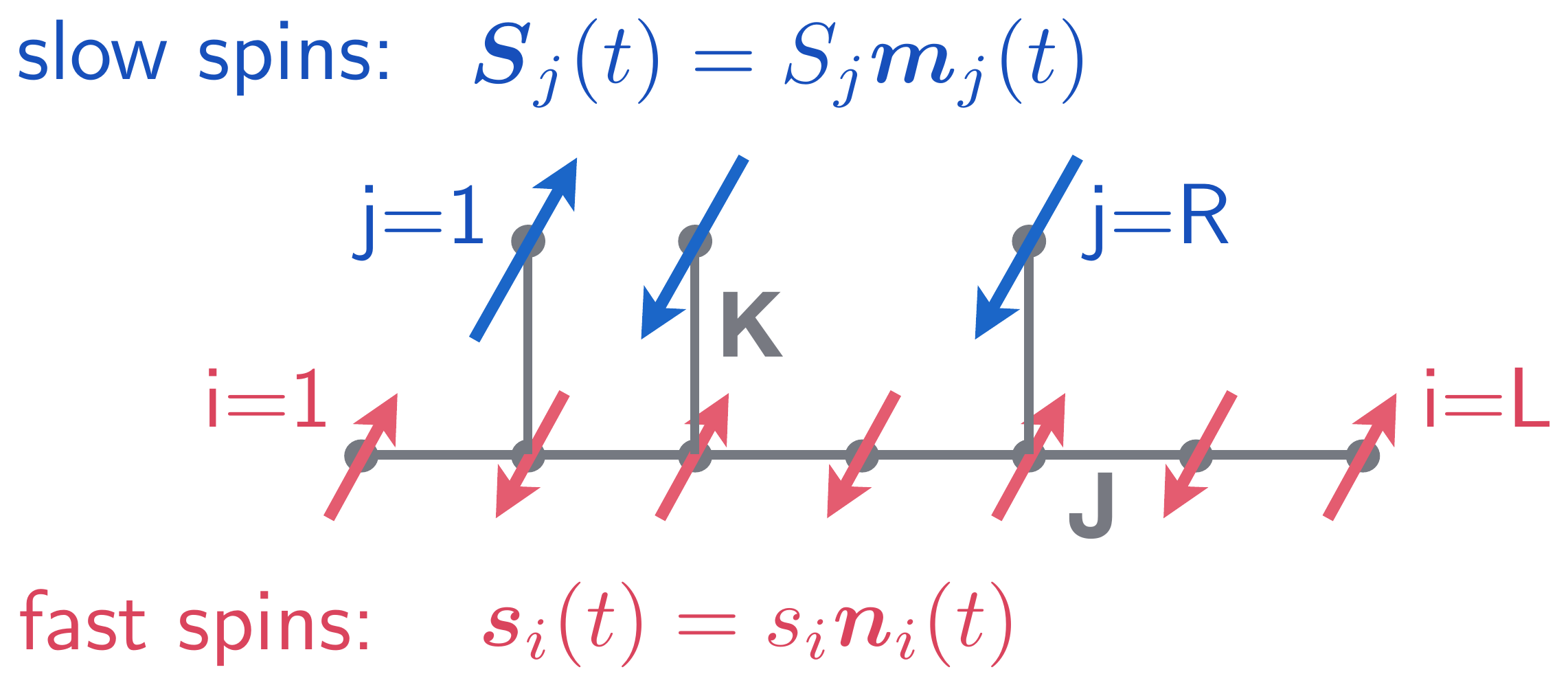}
\caption{
$L$ fast spins $\ff s_{i}$ ($i=1,...,L$) on a lattice, mutually interacting via a Heisenberg coupling $J$ and interacting via a local coupling $K$ with $R$ slow spins $\ff S_{j}$ ($j=1,...,R$). 
}
\label{fig:model}
\end{figure}

\paragraph{\color{blue} Classical spin system.}  

We consider an interacting system of classical spins, which can be divided into fast and slow spins. 
The $L$ fast spins $\ff s_{i} = s_{i} \ff n_{i}$ are assumed to be located on the sites $i=1,...,L$ of some lattice and interact via an isotropic Heisenberg exchange coupling $J_{ii'}$.
Here, $s_{i}$ is the length of the classical vector $\ff s_{i}$ and $\ff n_{i} = \ff s_{i} / s_{i}$ the corresponding unit vector. 
The fast spins are coupled to $R$ slow spins $\ff S_{j} = S_{j} \ff m_{j}$ via exchange couplings $K_{ij}$.
The slow unit vector at site $j$ is $\ff m_{j} = \ff S_{j}/S_{j}$.
The equations of motion,  
\be
  \dot{\ff s}_{i} = \frac{\partial H}{\partial \ff s_{i}} \times \ff s_{i}
  \: , \quad 
  \dot{\ff S}_{j} = \frac{\partial H}{\partial \ff S_{j}} \times \ff S_{j}
  \; , 
\label{eq:hameq}  
\ee
are obtained from the system's Hamiltonian
\be
H 
=
\frac{1}{2} \sum_{ii'} J_{ii'} \ff s_{i} \ff s_{i'}
+
\sum_{ij} K_{ij} \ff s_{i} \ff S_{j}
-
\sum_{j} \ff S_{j} \ff B
\: .
\label{eq:ham}
\ee
Fig.\ \ref{fig:model} sketches a possible realization with a one-dimensional lattice of fast spins and nearest-neighbor antiferromagnetic (AF) Heisenberg coupling $J$ and with local AF coupling $K$ to the slow spins. 
Here, the characteristic time scale of the fast-spin subsystem is given by $J^{-1}$ ($\hbar \equiv 1$), and the slow spins are subjected to an external magnetic field driving the slow-spin subsystem on a time scale $B^{-1}$.
Note that the equations of motion preserve the lengths of $\ff s_{i}$ and of $\ff S_{j}$ which  allows us to absorb constants, like gyromagnetic ratios, in $s_{i}$ and $S_{j}$.

The considered setup could mimic the magnetic properties of, e.g., magnetic atoms with magnetic moments $S_{j}$ on a magnetic solid surface \cite{Wie09}, or magnetic molecules \cite{FEH19}, etc.
The Hamiltonian could be extended by additional couplings between the slow spins or by anisotropic terms, and various alternative geometries are conceivable.
With Eq.\ (\ref{eq:ham}) we focus on a concrete Hamiltonian just to be specific, while our arguments are general.

The time evolution of an initial spin configuration is governed by the coupled nonlinear system of ordinary differential equations of motion (\ref{eq:hameq}).
It is typically exponentially sensitive to perturbations and quickly gets chaotic \cite{SJ05}. 
Here, our goal is to study a parameter regime, where the system exhibits two very different fast' and slow time scales and where the fast spins (almost) instantaneously follow the motion of the slow ones.
In this adiabatic limit, one can expect a strong conceptual simplification, providing us with an effective theory for the slow degrees of freedom only. 
As we will argue below, however, the slow-spin dynamics is additionally affected by {\em local} topological properties of the fast-spin subsystem, which give rise to unconventional effects.

\paragraph{\color{blue} Adiabatic limit.}  

The adiabatic limit is defined by a parameter range of the Hamiltonian where, at any instant of time $t$, the configuration of the fast spins $\ff s(t) \equiv (\ff s_{1}(t), ... , \ff s_{L}(t))$ is the ground-state configuration $\ff s(t) = \ff s_{0}(\ff S(t))$, for the present configuration $\ff S(t) \equiv (\ff S_{1}(t), ..., \ff S_{R}(t))$ of the slow spins given at the respective time $t$. 
Using fast and slow unit-vector configurations $\ff n(t) \equiv (..., \ff n_{i}(t), ...)$ and $\ff m(t) \equiv (..., \ff m_{j}(t), ...)$, respectively, we have
\be
   \ff n(t) = \ff n_{0}(\ff m(t)) \: .
\label{eq:con}
\ee
The (approximate) realization of the adiabatic limit and the question, in which parameter regime adiabatic spin dynamics is observed, will strongly depend on the specific system considered. 
Realizations for a simple toy model will be discussed below.
When approaching the adiabatic limit in parameter space, the fast-spin dynamics will be more and more constrained to the time-dependent hyper surfaces (\ref{eq:con}) in $\ff n$-space, i.e., in the product of Bloch spheres, $\ff n \in \prod_{i=1}^{L} \mathbb{S}^{2}$. 
In this limit we can employ Eq.\ (\ref{eq:con}) for a strongly simplified description of adiabatic spin dynamics (ASD). 

It is tempting to derive the slow-spin dynamics solely from the effective Hamiltonian 
$H(\ff s, \ff S) \mapsto H(s \ff n_{0}(\ff m),S \ff m) \equiv H_{\rm eff}(\ff m)$ that is obtained using the  constraints (\ref{eq:con}).
This {\em naive} ASD thus amounts to solving the $R$ remaining equations of motion $S_{j} \dot{\ff m}_{j} = \partial H_{\rm eff} / \partial \ff m_{j} \times \ff m_{j}$ for $\ff S$, while $\ff n(t)$ can be obtained from Eq.\ (\ref{eq:con}).
We will demonstrate below, however, that this may lead to incorrect results. 

\paragraph{\color{blue} Adiabatic spin dynamics.}  

A correct strategy towards ASD can be based on the action principle
$
\delta  \int dt \, L (\ff n, \dot{\ff n}, \ff m, \dot{\ff m}) = 0
$
with the Lagrangian \cite{BK90}
\be
L
=
\sum_{i} \ff A(\ff n_{i}) s_{i} \dot{\ff n}_{i}
+
\sum_{j} \ff A(\ff m_{j}) S_{j} \dot{\ff m}_{j}
-
H(\ff n, \ff m) \: .
\label{eq:lag}
\ee
Here, the function $\ff A(\ff r)$ must satisfy $\nab \times \ff A(\ff r)=-\ff r / r^{3}$ and can thus be interpreted as the vector potential of a unit magnetic (Dirac) monopole located at $\ff r = 0$. 
In the standard gauge \cite{Dir31}:
\be
  \ff A(\ff r) = - \frac{1}{r^{2}} \frac{\ff e_{z} \times \ff r}{1 + \ff e_{z} \ff r / r} \: .
\label{eq:dirac}  
\ee
The straightforward calculation [see the Supplemental Material (SM) \cite{SM}, section A] shows that the Lagrangian equations, $(d/dt) (\partial L / \partial \dot{\ff n}_{i}) = \partial L / \partial \ff n_{i}$ (and analogously for $\ff m_{j}$), are equivalent with the Hamiltonian Eqs.\ (\ref{eq:hameq}, \ref{eq:ham}).
This justifies the Lagrangian (\ref{eq:lag}). 
We note in passing that for classical spin dynamics, the Lagrangian and Hamiltonian formulation of the real-time dynamics are {\em not} related via a Legendre transformation since this is singular (see the SM \cite{SM}, Sec.\ B).

To describe spin dynamics in the adiabatic limit, i.e., spin dynamics constrained to the manifold specified by Eq.\ (\ref{eq:con}), one may employ the action principle.
To this end, the holonomic constraints (\ref{eq:con}) are used to reduce the number of ``generalized coordinates'' and to define an effective Lagrangian for the slow-spin degrees of freedom only: 
$L_{\rm eff}(\ff m,\dot{\ff m}) \equiv L(\ff n_{0}(\ff m),(d/dt)\ff n_{0}(\ff m),\ff m,\dot{\ff m})$.
The ASD equations of motion for the slow spins $\ff m_{j}$ are then obtained from $\delta \int dt \, L_{\rm eff} = 0$, where $\delta$ indicates variation of the slow-spin configuration $\ff m$ only.
The calculation is completely straightforward, as detailed in Sec.\ C of the SM \cite{SM}, and results in 
\be
S_{j} \dot{\ff m}_{j} = \frac{\partial H_{\rm eff}}{\partial \ff m_{j}} \times \ff m_{j} + \ff T_{j} \times \ff m_{j} \, . 
\label{eq:asd}
\ee
This is the central result of our work.

As compared to the ``naive'' adiabatic theory, there is an additional term due to a field
\be
  \ff T_{k} =  \ff T_{k}(\ff m,\dot{\ff m}) = \sum_{l,\mu\nu} \Omega_{k\mu,l\nu}(\ff m) \dot{m}_{l\nu} \ff e_{\mu} \: ,
\label{eq:tj}  
\ee
with $\Omega_{k\mu,l\nu}(\ff m) = 4\pi \sum_{i} s_{i} e^{(i)}_{k\mu,l\nu}(\ff m)$, and where
\be
e^{(i)}_{k\mu,l\nu}(\ff m)
=
\frac{1}{4\pi}
\frac{\partial \ff n_{0,i}(\ff m) }{\partial m_{k\mu}}
\times
\frac{\partial \ff n_{0,i}(\ff m) }{\partial m_{l\nu}}
\cdot
\ff n_{0,i}(\ff m) 
\: 
\label{eq:cd}
\ee
is a rank-2 tensor for each pair of sites $k,l$ and antisymmetric: $e^{(i)}_{k\mu,l\nu}(\ff m) = - e^{(i)}_{l\nu,k\mu}(\ff m)$.
The tensor describes properties of the fast-spin subsystem only and is thus not (directly) affected by a finite magnetic field $\ff B$ coupling to the slow spins in the Hamiltonian (\ref{eq:ham}).
Under time reversal (TR), $e^{(i)}_{k\mu,l\nu}(\ff m) \mapsto e^{(i)}_{k\mu,l\nu}(-\ff m) = - e^{(i)}_{k\mu,l\nu}(\ff m)$, since $\ff n_{0,i}(-\ff m) = - \ff n_{0,i}(\ff m)$ (independent of $\ff B$).
Note that this implies $\ff T_{j} \mapsto - \ff T_{j}$ under TR, and that the form of the equation of motion (\ref{eq:asd}) is TR invariant.

\paragraph{\color{blue} Topological spin torque.}  
Each tensor element for fixed $k,l$ defines a topological charge {\em density}, which becomes a quantized homotopy invariant when integrated.
This is reminiscent of the skyrmion density \cite{Bra12,FBT+16,SHP+17} but for skyrmions living on a product of Bloch spheres rather than in Euclidean space.
We use arguments analogous to those invoked for demonstrating the topological quantization of the spin Hall effect \cite{TKNN82,QWZ06}:
The configuration space of the slow spins at sites $k$ and $l$ ($k\ne l$) is given by $\mathbb{S}^{2} \times \mathbb{S}^{2}$. 
If $\ff m_{k}$ and $\ff m_{l}$ perform a cyclic motion in this space, i.e., cover a subspace $\Sigma \cong \mathbb{S}^{1} \times \mathbb{S}^{1} \cong \mathbb{T}^{2}$, then $4\pi e^{(i)}_{k\mu,l\nu}(\ff m)$ is the Jacobian of the map 
$\mathbb{T}^{2} \to \mathbb{S}^{2}$, $(\ff m_{k}, \ff m_{l}) \mapsto \ff n_{i}$, where $\mathbb{S}^{2}$ is the configuration space of the fast spin at site $i$.
The same holds for $k=l$, where the subspace is $\Sigma \cong \mathbb{S}^{2}$.
Therefore, integrating the Jacobian over $\Sigma$, just gives the total area of the image of $\Sigma$ on $\mathbb{S}^{2}$, divided by $4\pi$.
This is a topological winding number $e^{(i)}_{kl}$ with a quantized value $e^{(i)}_{kl} \in \mathbb{Z}$ indicating how many times the Bloch sphere of $\ff n_{i}$ is covered by the image.
Under TR $e^{(i)}_{kl} \mapsto - e^{(i)}_{kl}$.

The adiabatic equations of motion (\ref{eq:asd}) tell us that already the {\em local} topological properties of the fast spins, i.e., the topological charge {\em densities} $e^{(i)}_{k\mu,l\nu}(\ff m)$ of the fast-spin subsystem, play a decisive role for the slow-spin dynamics. 
The resulting topological spin torque $\ff T_{j} \times \ff m_{j}$ is largely independent of microscopic details, such as coupling strengths, but depends on geometrical system properties.
Although the topological torque involves the time derivative $\dot{\ff m}_{l}$, see Eq.\ (\ref{eq:tj}), it respects total-energy conservation, unlike a Gilbert-damping \cite{Gil55,Gil04} or an anti-damping term \cite{KSF+14,FBM14}, see the SM \cite{SM}, Sec.\ D.

\begin{figure}[t]
\includegraphics[width=0.95\columnwidth]{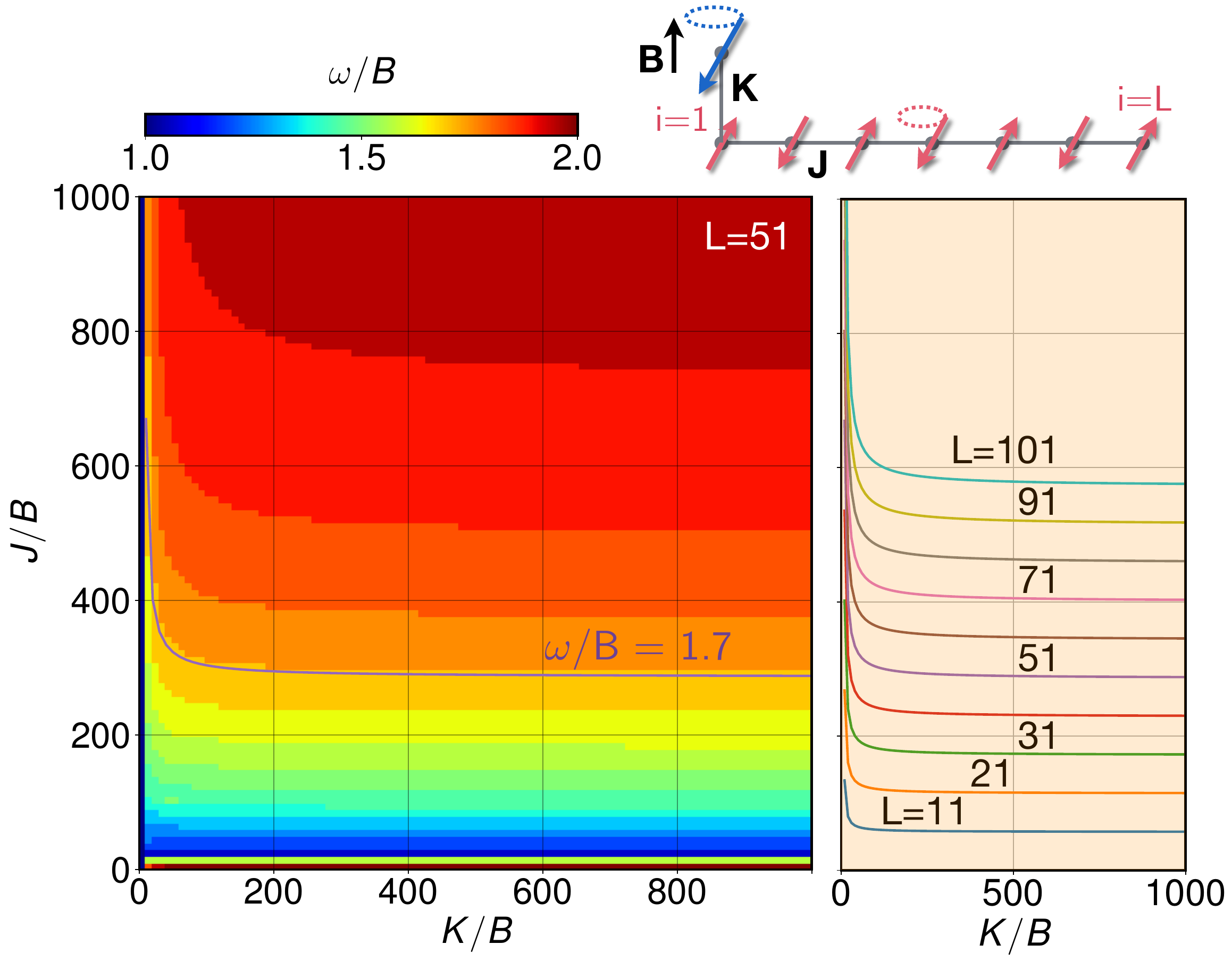}
\caption{
Precession frequency (see colorbar, top panel) for a single slow impurity spin ($S=1$) driven by a magnetic field of strength $B$ as a function of the local AF exchange coupling $K$ and of the AF nearest-neighbor Heisenberg coupling $J$.
Results as obtained numerically from Eq.\ (\ref{eq:hameq}). 
Initially, the system is in its ground state with $\ff S \propto \ff e_{x}$. 
At time $t=0$, the field $\ff B=B \ff e_{z}$ is suddenly switched on.
{\em Dark red:} anomalous precession frequency $\omega/B \to 2$ as predicted to result from the topological spin torque, indicating the realization of the adiabatic limit. 
{\em Dark blue:} standard Larmor frequency $\omega/B \to 1$. 
{\em Left:} Results for a linear chain with $L=51$ (fast) spins $s=1/2$.
The blue line is a fit to the data indicating a frequency ($\omega / B=1.7$) already close to ASD limit value.
{\em Right:} evolution of the anomalous regime, tentatively defined as $\omega/B > 1.7$ with system size $L$.
}
\label{fig:prec}
\end{figure}

\paragraph{\color{blue} Single slow spin.} 
The $R=1$ case allows us to compute the topological spin torque analytically. 
We consider a single slow spin $\ff S=S\ff m$, driven by a field $\ff B$ and coupled via a local antiferromagnetic (AF) exchange $K>0$ to the first spin ($i=1$) of an open one-dimensional array of $L$ fast spins $\ff s_{i} = s \ff n_{i}$ (with constant $s_{i}=s$), mutually interacting by AF nearest-neighbor Heisenberg couplings of strength $J>0$ (see inset in Fig.\ \ref{fig:prec}).

For $R=1$ the $k=l=1$ element of the topological charge density defines a pseudo-vector field $\ff e^{(i)}(\ff m) = \frac12 \sum_{\mu\nu} \varepsilon_{\mu\nu\rho} \, e^{(i)}_{1\mu,1\nu}(\ff m) \, \ff e_{\rho}$.
Note that this can be seen as a magnetic vorticity \cite{Coo99}, but in $\ff m$-space. 
At the same time $4\pi s \sum_{i} \ff e^{(i)}(\ff m)$ can be interpreted as a ``magnetic field'' and $\ff T$ as the corresponding ``Lorentz force'' in $\ff m$-space, as is seen from Eqs.\ (\ref{eq:asd}) and (\ref{eq:tj}).

The naive expectation is $S \dot{\ff m} = \partial H_{\rm eff} / \partial \ff m \times \ff m = S\ff m \times \ff B$, i.e., the spin precesses with Larmor frequency $\omega_{L} = B$ around the axis defined by the direction of the physical field.
Note that here the fast spins do not contribute a torque since their ground state is a classical N\'eel state aligned to $\ff m$, i.e., $\ff n_{0,i} = (-1)^{i} \ff m$.

To get the correct equation of motion, we first compute the topological charge density. 
Eq.\ (\ref{eq:cd}) yields $e^{(i)}_{\mu\nu}(\ff m) = (-1)^{i} \ff e_{\mu} \times \ff e_{\nu} \cdot \ff m / 4\pi$.
Using spherical coordinates, one finds $e^{(i)} = (-1)^{i} \iint d\vartheta d\varphi \:(\partial \ff m/\partial \vartheta) \times (\partial \ff m/\partial \varphi) \cdot \ff m = (-1)^{i} k$ with $k\in \mathbb{Z}$ and independent of $i$. 
With $e^{(i)}_{\mu \nu}(\ff m)$ at hand, Eq.\ (\ref{eq:tj}) provides us with the field $\ff T$: 
While $\ff T=0$, if $L$ is even, we get $\ff T= s \ff m \times \dot{\ff m}$ for odd $L$, i.e., when there is a finite total spin of the fast-spin subsystem. 
While even-odd effects in the ground-state magnetic structure are well known (see e.g.\ \cite{LDB08}),
the topological spin torque leads to an even-odd effect in the spin {\em dynamics}.

For odd $L$ the anomalous spin torque takes the simple form $\ff T \times \ff m = s \dot{\ff m}$, such that the full adiabatic equation of motion reads
$S \dot{\ff m} = S \ff m \times \ff B + s \dot{\ff m}$.
Combining the two $\dot{\ff m}$-dependent terms, this has precisely the form of the Landau-Lifschitz equation, $\dot{\ff S} = \ff S \times \widetilde{\ff B}$, but with a renormalized field strength $\widetilde{B}$.
Hence we again find a simple precessional motion of the slow spin albeit with renormalized precession frequency
\be
  \omega = \frac{1}{1-s/S} \cdot \omega_{L} \: ,
\label{eq:ano}  
\ee
which is higher than the Larmor frequency $\omega_{L} = B$ for $S>s$, e.g., $\omega=2B$ for $S=1$ and $s=1/2$, while for $S<s$, the orientation of the precession is inverted. 

We conclude that, already for the $R=1$ case, there are nontrivial effects of the anomalous spin torque. 
Furthermore, the specialization to $R=1$ is instructive as this allows for comparison with a system, where the fast spins are replaced by conduction electrons, see Ref.\ \cite {SP17}:
For the quantum-classical system, the same renormalization of the precessional motion has been found, but the role of the topological charge density $e^{(i)}_{\mu\nu}(\ff m)$ is played by the (spin) Berry curvature.
This underpins the large independency of the novel spin torque from microscopic details of the fast subsystem.

\paragraph{\color{blue} Realization of the adiabatic limit.}  
An important question concerns the conditions for which the constraints (\ref{eq:con}) are (approximately) satisfied.
Previous work on spin dynamics in the $s$-$d$ impurity model \cite{SP15,SRP16a,SRP16b,BN19} has shown that the real-time evolution of the classical spin is {\em almost} adiabatic in large regions of parameter space. 
It is thus tempting to expect almost adiabatic motion in a certain parameter range and hence an anomalous precession frequency, as predicted by Eq.\ (\ref{eq:ano}), for the present case of a purely classical spin system as well. 
Note that for $s=S$ in Eq.\ (\ref{eq:ano}), $\omega$ diverges, indicating that adiabatic motion is not possible in this case.

We have checked the predictions of ASD by numerically solving the {\em full} equations of motion (\ref{eq:hameq}) using a high-order Runge-Kutta technique for a slow spin of length $S=1$.
As expected, for a system with even $L$, the observed precession frequency is very close to the standard Larmor frequency, i.e., $\omega \approx \omega_{L} = B$, including those parameter regimes where the fast spins almost instantaneously follow the slow-spin dynamics, but with one exception discussed below.
For systems with odd $L$, on the other hand, we in fact find an anomalous frequency $\omega \approx 2 \omega_{L}$ resulting from the topological spin torque, see Eq.\ (\ref{eq:cd}), 
for model parameters where the dynamics is adiabatic. 
As is seen in Fig.\ \ref{fig:prec}, this is the case whenever the field strength $B$ is small compared to $K$ and $J$. 

In the simple two-spin system for $L=1$ or, equivalently, for $J=0$, we have $\omega \approx 2 \omega_{L}$ if the two spins are strongly coupled, i.e., for $K/B \gg 1$.
An additional weak coupling to $L>1$ fast spins, i.e., $J/K \ll 1$, then only slightly perturbs the tightly bound two-spin subsystem and the corresponding anomalous dynamics.
This explains the increase of $\omega$ with {\em decreasing} $J$ and the narrow red-colored anomalous range visible in Fig.\ \ref{fig:prec} for extremely small $J/B$.
Since this effect results from the proximity to the $L=1$ case, it also shows up for {\em even} L (not shown).

With increasing system size, the characteristic time scale $1/B$ of the slow spin must increase as well to keep the dynamics in the adiabatic limit.
This is obvious as the information on the state of the impurity spin must propagate through the whole system to allow the fast spins to align. 
Assuming that the propagation time grows at least linearly with system size $L$, we must have $1/B \sim L/J$ for large $K/B$. 
We find that the numerical results can be fitted nicely assuming $1/B \propto 1/K + (L-1)/J$ (see the line in the main panel of Fig.\ \ref{fig:prec}). 
The right panel demonstrates the linear growth of the necessary time scale with $L$.

This implies that, for the model considered here, nonadiabatic dynamics must be expected in the thermodynamic limit $L\to \infty$.
At the same time the observed even-odd effect becomes irrelevant and we must generalize the theory to mixed states.
Hence, there is a big {\em de facto} but also a big conceptual difference compared to quantum-classical systems \cite{SP15,SRP16a,SP17}, where the adiabatic limit can be controlled by the gap in the electronic structure and the adiabatic theorem \cite{Mes61}.

\paragraph{\color{blue} Conclusions.}  
Classical spin systems are frequently employed in various contexts, 
such as atomistic spin dynamics of condensed-matter, nanostructured and molecular systems. 
Here, we have considered prototypical classical Heisenberg-like models and have demonstrated that time-scale separation generally leads to the emergence of a topological spin torque.
This may profoundly affect the dynamics as could be seen already for a simplistic toy model. 
The full implications of the adiabatic spin-dynamics theory, however, are yet to be worked out.

Promising applications include systems of magnetic atoms indirectly coupled via a magnetic host. 
In the weak exchange-coupling case, we expect slow RKKY-like dynamics in nanostructures \cite{YHV18} 
or molecules \cite{FEH19} but with additional impact due to the topological spin torque. 
Time-scale separation is also very much triggered by anisotropic interactions, such that surfaces and interfaces but also nanoislands \cite{BRKW05}
and chains \cite{LDB08} of magnetic systems should be very sensitive to topological effects in the real-time domain. 
Of particular interest could be the (weak) coupling of magnetic moments to materials with an electronic or spin structure that is intrinsically topological. 
An intriguing question is how ASD can couple to or even uncover topological properties of the host.
This may be addressed, e.g., by studying spins weakly coupled to a correlated topological lattice-fermion model, such as the Kane-Mele-Hubbard \cite{KM} or related spin-only models \cite{RLH10,AMM12}, or spin systems with topological magnon excitation spectra \cite{MU19}.

\acknowledgments
This work was supported by the Deutsche Forschungsgemeinschaft (DFG) through the Cluster of Excellence ``Advanced Imaging of Matter'' - EXC 2056 - project ID 390715994, and by the DFG 
Sonderforschungsbereich 925 ``Light-induced dynamics and control of correlated quantum systems''
(project B5).

%

\begin{widetext}
\newpage
\mbox{}
\setcounter{page}{1}

\begin{center}
{\large \bfseries 
Topological spin torque emerging in classical-spin systems with different time scales

\mbox{}\\

--- Supplemental Material ---
}

\mbox{}\\

{
Michael Elbracht$^{1}$, 
Simon Michel$^{1}$, 
and
Michael Potthoff$^{1,2}$
}

\mbox{}\\[-2mm]

{
\small \it 
$^{1}$I. Institute of Theoretical Physics, Department of Physics, 

University of Hamburg, Jungiusstra\ss{}e 9, 20355 Hamburg, Germany

$^{2}$The Hamburg Centre for Ultrafast Imaging, Luruper Chaussee 149, 22761 Hamburg, Germany
}

\end{center}


\paragraph{\color{blue} 
Section A: Lagrangian equations of motion.} 

Consider a Lagrangian of the form
\be
L = \sum_{i} \ff A(\ff n_{i}) s_{i} \dot{\ff n}_{i}
-
H(\ff n) 
- 
\sum_{i}
\lambda_{i} (\ff n_{i}^{2} - 1)
\: , 
\label{eq:lag1}
\ee
as given by Eqs.\ (\ref{eq:lag}) and (\ref{eq:dirac}) but, for simplicity, for a single type of spins only. 
We have $\ff s= (\ff s_{1}, ..., \ff s_{L})$, $\ff s_{i} = s_{i} \ff n_{i}$, $H({\ff n}) \equiv H(\ff s)$ and we have explicitly added Lagrange-multiplier terms $\propto \lambda_{i}$ to keep track of the constraints $\ff n_{i}^{2}=1$.
The spin directions $\ff n_{i}(t)$ and the Lagrange parameters $\lambda_{i}(t)$ are obtained from the condition that the action corresponding to $L$ be stationary.
The Lagrangian equations of motion, derived from the action principle, read 
\be
0 = \frac{d}{dt} \frac{\partial L}{\partial \dot{n}_{i\alpha}} - \frac{\partial L}{\partial n_{i\alpha}}
 = s_{i} \sum_{\beta} \left(
 \frac{\partial A_{\alpha}(\ff n_{i})}{\partial n_{i\beta}} 
 -
 \frac{\partial A_{\beta}(\ff n_{i})}{\partial n_{i\alpha}} 
 \right)
 \dot{n}_{i\beta}
 + \frac{\partial H(\ff n)}{\partial n_{i\alpha}}
 + 2 \lambda_{i} n_{i\alpha} \: ,
\ee
with $\alpha=x,y,z$.
Using the vector notation, we get
\be 
 - s_{i} \dot{\ff n}_{i} \times (\nab \times \ff A(\ff n_{i}))
 + \frac{\partial H(\ff n)}{\partial \ff n_{i}} + 2 \lambda_{i} \ff n_{i}
 = 0
 \: .
\ee
Inserting the curl of the vector potential, $\nab \times \ff A(\ff n_{i})=-\ff n_{i} / n_{i}^{3}$, yields:
\be
 s_{i} \dot{\ff n}_{i} \times \ff n_{i} / n_{i}^{3}
 + \frac{\partial H(\ff n)}{\partial \ff n_{i}} + 2 \lambda_{i} \ff n_{i}
 =0 
 \: .
\label{eq:der}
\ee
Taking the cross product from the right, $\times \ff n_{i}$, as well as the dot product $\cdot \ff n_{i}$, provides us with the following two equations:
\be
 s_{i} (\dot{\ff n}_{i} \times \ff n_{i}) \times \ff n_{i} / n_{i}^{3}
 + \frac{\partial H(\ff n)}{\partial \ff n_{i}} \times \ff n_{i}  = 0
 \; , \qquad
 \frac{\partial H(\ff n)}{\partial \ff n_{i}} \ff n_{i} + 2 \lambda_{i} \ff n_{i}^{2} = 0
\ee
which are equivalent with Eq.\ (\ref{eq:der}). 

The constraint $\ff n_{i}^{2}=1$ and the second equation fix the Lagrange parameter as: $\lambda_{i} = - (\partial H(\ff n) / \partial \ff n_{i}) \ff n_{i} / 2$.
Using the constraint to simplify the first equation, results in:
\be
 s_{i} \dot{\ff n}_{i} 
 = \frac{\partial H(\ff n)}{\partial \ff n_{i}} \times \ff n_{i}  
\ee
or, equivalently, 
\be
 \dot{\ff s}_{i} 
 = \frac{\partial H(\ff s)}{\partial \ff s_{i}} \times \ff s_{i} \: , 
\ee
i.e., the Hamilton equations of motion for $\ff s$, see Eq.\ (\ref{eq:hameq}).
\\

\paragraph{\color{blue} 
Section B: Hamiltonian vs.\ Lagrangian spin dynamics.} 

One way to see that the Hamiltonian and the Lagrangian formulation of spin dynamics are not related via a Legendre transformation is the following: 
Starting from a generic Hamilton function for classical spins $\ff s_{i}$, e.g.,
\be
  H(\ff s) = \frac{1}{2} \sum_{ij} J_{ij} \ff s_{i} \ff s_{j} - \sum_{i} \ff B_{i} \ff s_{i} \: , 
  \label{eq:sham}
\ee
and introducing coordinates $\ff q_{i}$ and momenta $\ff p_{i}$ such that
$\ff s_{i} = \ff q_{i} \times \ff p_{i}$, one gets the Hamiltonian
\be
  H(\ff q, \ff p) = H(\ff s)\big|_{\ff s = \ff q \times \ff p}
\ee
and the resulting canonical equations of motion for $(\ff q_{i}, \ff p_{i})$:
\be
  \dot{\ff q}_{i} 
  =
  \frac{\partial H(\ff q, \ff p)}{\partial \ff p_{i}} 
  = 
  \frac{\partial H(\ff s)}{\partial \ff s_{i}} \Big|_{\ff s = \ff q \times \ff p}
  \times \ff q_{i}
  \: , \qquad
  \dot{\ff p}_{i} 
  =
  - \frac{\partial H(\ff q, \ff p)}{\partial \ff q_{i}} 
  = 
  \frac{\partial H(\ff s)}{\partial \ff s_{i}} \Big|_{\ff s = \ff q \times \ff p}
  \times \ff p_{i}
  \: .
\label{eq:hamq} 
\ee
One easily verifies: $\dot{\ff s_{i}} = \ff q_{i} \times \dot{\ff p}_{i} + \dot{\ff q}_{i}\times \ff p_{i} = (\partial H(\ff s)/\partial \ff s_{i}) \times \ff s_{i}$.
Now, it is tempting to define a Lagrangian via 
\be
L (\ff q,\dot{\ff q}) = \sum_{i} \ff p_{i}(\ff q,\dot{\ff q}) \dot{\ff q}_{i} - H(\ff q, \ff p(\ff q,\dot{\ff q}))
\: , 
\ee
where the function $\ff p(\ff q,\dot{\ff q})$ would be obtained by solving Eq.\ (\ref{eq:hamq}) for $\ff p_{i}$.
However, for a Hamiltonian as given by Eq.\ (\ref{eq:sham}), the equations,
\be
\dot{\ff q}_{i} 
=
\frac{\partial H(\ff s)}{\partial \ff s_{i}} \Big|_{\ff s = \ff q \times \ff p} \times \ff q_{i}
=
\sum_{j} J_{ij} (\ff q_{j} \times \ff p_{j}) \times \ff q_{i} - \ff B_{i} \times \ff q_{i}
\: ,
\ee  
form an inhomogeneous linear system of equations $\underline{A} \cdot \ff p = \ff b$ for the unknowns $\ff p_{i}$, which is necessarily singular as the coefficient matrix $\underline{A}$ with elements ($\alpha,\beta \in \{x,y,z\}$),
\be
A_{i\alpha,j\beta} 
= 
J_{ij} (\delta_{\alpha \beta} \sum_{\gamma} q_{i\gamma} q_{j\beta} - q_{j\alpha} q_{i\beta})
  \: ,
\ee
is singular since $\underline{A} \cdot \ff q = 0$, irrespective of the interaction parameters $J_{ij}$.
\\

\paragraph{\color{blue} 
Section C: ASD equation of motion.} 

The slow-spin dynamics is derived from the effective Lagrangian 
\be
L_{\rm eff}(\ff m,\dot{\ff m}) \equiv L(\ff n_{0}(\ff m),(d/dt)\ff n_{0}(\ff m),\ff m,\dot{\ff m}) \: .
\ee
With
\be
\frac{d}{dt} \ff n_{0,i}(\ff m) = \sum_{j} (\dot{\ff m}_{j} \nab_{j}) \ff n_{0,i}(\ff m)
\ee
we find:
\ba
L_{\rm eff}(\ff m, \dot{\ff m})
=
\sum_{j} \ff A(\ff m_{j}) S_{j} \dot{\ff m}_{j}
+
\sum_{i} \ff A(\ff n_{0,i}(\ff m)) s_{i} 
\sum_{j} (\dot{\ff m}_{j} \nab_{j}) \ff n_{0,i}(\ff m)
-
H_{\rm eff}(\ff m) \; , 
\label{eq:leff}
\ea
where $i=1,...,L$ and $j=1,...,R$, and where $H_{\rm eff}(\ff m) = H(s \ff n_{0}(\ff m),S \ff m)$.
To get the Lagrange equations of motion, we first compute
\ba
\frac{\partial}{\partial \ff m_{k}} L_{\rm eff}(\ff m, \dot{\ff m})
&=&
\sum_{\beta} S_{k}
\nab_{k} A_{\beta}(\ff m_{k}) \dot{m}_{k\beta}
+ \sum_{i\beta} s_{i}  A_{\beta}(\ff n_{0,i}(\ff m))
\sum_{j} (\dot{\ff m}_{j} \nab_{j}) \nab_{k} n_{0,i\beta}(\ff m)
\nonumber \\
&+& \sum_{i\alpha\beta} s_{i} 
\frac{\partial A_{\beta}(\ff n_{0,i}(\ff m))}{\partial n_{0,i\alpha}}
\nab_{k} n_{0,i\alpha}(\ff m)
\sum_{j} (\dot{\ff m}_{j} \nab_{j})  n_{0,i\beta}(\ff m)
-
\nab_{k} H_{\rm eff}(\ff m) \; .
\label{eq:dldm}
\ea
Here, $\nab_{j} = \partial / \partial \ff m_{j}$, and Greek indices $\alpha, \beta, ... \in \{x,y,z\}$. 
Next, 
\ba
\frac{\partial}{\partial \dot{\ff m}_{k}} L_{\rm eff}(\ff m, \dot{\ff m})
=
S_{k} \ff A(\ff m_{k}) 
+
\sum_{i\alpha} s_{i} A_{\alpha}(\ff n_{0,i}(\ff m))   
\nab_{k} n_{0,i\alpha}(\ff m) 
\; ,
\label{eq:dldmd}
\ea
which yields
\ba
\frac{d}{dt} \frac{\partial L_{\rm eff}(\ff m, \dot{\ff m})}{\partial \dot{\ff m}_{k}} 
&=&
\sum_{\alpha} S_{k}
(\nab_{k} A_{\alpha}(\ff m_{k}) \dot{\ff m}_{k}) \ff e_{\alpha}
+ \sum_{ij\alpha\beta} s_{i} \frac{\partial A_{\alpha}(\ff n_{0,i}(\ff m))}{\partial n_{0,i\beta}}
(\dot{\ff m}_{j} \nab_{j} n_{0,i\beta}(\ff m) )
\nab_{k} n_{0,i\alpha}(\ff m)
\nonumber \\
&+&
\sum_{i\alpha} s_{i} A_{\alpha}(\ff n_{0,i}(\ff m) 
\sum_{j} \nab_{k}
\left(
\dot{\ff m}_{j} \nab_{j} n_{0,i\alpha}(\ff m) 
\right)
\; .
\ea
The last term equals the second term on the right-hand side of Eq.\ (\ref{eq:dldm}) in the Lagrange equations, since $\nab_{k}$ and $\dot{\ff m}_{j} \nab_{j}$ commute, such that we are left with:
\ba
0 &=& \frac{d}{dt} \frac{\partial L_{\rm eff}}{\partial \dot{\ff m}_{k}} 
-
\frac{\partial L_{\rm eff}}{\partial {\ff m}_{k}} 
=
\sum_{\alpha} S_{k}
\dot{\ff m}_{k} \nab_{k} A_{\alpha}(\ff m_{k}) \ff e_{\alpha}
- \sum_{\beta} S_{k}
\dot{m}_{k\beta} \nab_{k} A_{\beta}(\ff m_{k}) 
+
\nab_{k} H_{\rm eff}(\ff m) 
\nonumber \\
&+& \sum_{ij\alpha\beta} s_{i} \frac{\partial A_{\alpha}(\ff n_{0,i}(\ff m))}{\partial n_{0,i\beta}}
(\dot{\ff m}_{j} \nab_{j} n_{0,i\beta}(\ff m))
\nab_{k} n_{0,i\alpha}(\ff m)
- \sum_{ij\alpha\beta} s_{i} 
\frac{\partial A_{\beta}(\ff n_{0,i}(\ff m))}{\partial n_{0,i\alpha}}
\nab_{k} n_{0,i\alpha}(\ff m)
(\dot{\ff m}_{j} \nab_{j}) \ff n_{0,i\beta}(\ff m)
\nonumber \\
&=&
S_{k} (\nab_{k} \times \ff A(\ff m_{k})) \times \dot{\ff m}_{k} 
+
\nab_{k} H_{\rm eff}(\ff m) 
+ \ff T_{k}
\label{eq:eomtk}
\: ,
\ea
where $\ff T_{k}$ stands for the last two terms. 
Taking the cross product from the right, $(...)\times \ff m_{k}$, we find
\ba
S_{k} ((\nab_{k} \times \ff A(\ff m_{k})) \times \dot{\ff m}_{k} ) \times \ff m_{k} 
+
\nab_{k} H_{\rm eff}(\ff m) \times \ff m_{k}
+ \ff T_{k} \times \ff m_{k}
= 0
\: .
\label{eq:tk}
\ea
Using $\nab_{k} \times \ff A(\ff m_{k}) = - \ff m_{k} / m_{k}^{3}$, expanding the remaining double cross product and exploiting that $\ff m_{k}$ is a unit vector, yields:
\be
S_{k} \dot{\ff m}_{k}
=
\nab_{k} H_{\rm eff}(\ff m) \times \ff m_{k}
+ \ff T_{k} \times \ff m_{k} \: , 
\ee
which, apart from the extra term involving $\ff T_{k}$, just recovers the ``naive'' adiabatic spin dynamics.

Note that actually we should have added Lagrange-multiplier terms, $L_{\rm eff}(\ff m, \dot{\ff m}) \mapsto L_{\rm eff}(\ff m, \dot{\ff m}) - \sum_{k} \lambda_{k} (\ff m_{k}^{2}-1)$, to account for the normalization conditions $\ff m_{k}^{2}=1$.
However, this would have resulted in an additional summand $2\lambda_{k} \ff m_{k}$ on the r.h.s.\ of Eq.\ (\ref{eq:eomtk}) only, which does not contribute after taking the cross product $(...) \times \ff m_{k}$.
On the other hand, taking the dot product, $(...)\cdot \ff m_{k}$, of Eq.\ (\ref{eq:eomtk}), just yields the necessary conditional equation for $\lambda_{k}$, if this was required.

$\ff T_{k}$, if nonzero, gives rise to an additional spin torque $\ff T_{k} \times \ff m_{k}$.
From Eq.\ (\ref{eq:tk}) we can read off:
\be
\ff T_{k}
=
\sum_{ij\alpha\beta} s_{i} 
\left( 
\frac{\partial A_{\alpha}(\ff n_{0,i}(\ff m))}{\partial n_{0,i\beta}} 
-
\frac{\partial A_{\beta}(\ff n_{0,i}(\ff m))}{\partial n_{0,i\alpha}}
\right)
(\dot{\ff m}_{j} \nab_{j}) n_{0,i\beta}(\ff m) 
\,
\nab_{k} n_{0,i\alpha}(\ff m) \: .
\ee
Exploiting once more the defining property of the vector potential, $\nab_{k} \times \ff A(\ff m_{k}) = - \ff m_{k} / m_{k}^{3}$, and using the normalization $m_{j}=1$ in the end, we find:
\ba
\ff T_{k}
&=&
\sum_{ij\alpha\beta\gamma} s_{i} 
\epsilon_{\alpha\beta\gamma}
\nab_{k} n_{0,i\alpha}(\ff m) \,
(\dot{\ff m}_{j} \nab_{j}) n_{0,i\beta}(\ff m) \, 
n_{0,i\gamma}(\ff m) 
\nonumber \\
&=&
\sum_{i} s_{i} \sum_{l} \sum_{\mu\nu}
\nabla_{k\mu} \ff n_{0,i}(\ff m) 
\times
\nabla_{l\nu}   \ff n_{0,i}(\ff m) 
\cdot
\ff n_{0,i}(\ff m) 
\:\dot{m}_{l\nu} \, \ff e_{\mu}
\ea
The scalar triple product defines an antisymmetric tensor of rank two:
\be
\Omega_{k\mu,l\nu}
=
\sum_{i} s_{i}\,
\frac{\partial \ff n_{0,i}(\ff m) }{\partial m_{k\mu}}
\times
\frac{\partial \ff n_{0,i}(\ff m) }{\partial m_{l\nu}}
\cdot
\ff n_{0,i}(\ff m) 
=
- \Omega_{l\nu,k\mu} \: . 
\ee
Hence:
\be
   \ff T_{k}
   =
  \sum_{l} \sum_{\mu\nu}
   \Omega_{k\mu,l\nu}
   \:
   \dot{m}_{l\nu} \, \ff e_{\mu}
   \: .
\ee
Note the following sum rule:
\be
   \sum_{k} \ff T_{k} \dot{\ff m}_{k}    
   = 0
   \: .
\label{eq:sumrule}
\ee

\paragraph{\color{blue} 
Section D: Total-energy conservation.} 

As the constraints (\ref{eq:con}) are time-independent and holonomic, total-energy conservation within the effective adiabatic theory is actually ensured by the general Lagrange formalism but can also be verified explicitly by computing the time derivative of the total energy:
\be
  \frac{dE}{dt} 
  =
  \frac{dH_{\rm eff}(\ff m)}{dt} 
  =
  \frac{d}{dt} \sum_{j} \ff A(\ff m_{j}) S_{j} \dot{\ff m}_{j}
  +
  \frac{d}{dt} \sum_{i} \ff A(\ff n_{0}(\ff m)) s_{i} 
  -
  \frac{d}{dt} L_{\rm eff}(\ff m, \dot{\ff m})
   \; .
\label{eq:econs}   
\ee
Using the Lagranian equations of motion, 
$(d/dt) (\partial L_{\rm eff} / \partial \dot{\ff m}_{k}) = \partial L_{\rm eff} / \partial {\ff m}_{k}$,
and Eq.\ (\ref{eq:dldmd}), we have
\ba
  \frac{d}{dt} L_{\rm eff}(\ff m, \dot{\ff m}) 
  &=&
  \sum_{j} \frac{\partial L_{\rm eff}}{\partial \ff m_{j}} \dot{\ff m}_{j}
  +
  \sum_{j} \frac{\partial L_{\rm eff}}{\partial \dot{\ff m}_{j}} \ddot{\ff m}_{j}
  =
  \sum_{j} \left( \frac{d}{dt} \frac{\partial L_{\rm eff}}{\partial \dot{\ff m}_{j}} \right) \dot{\ff m}_{j}
  +
  \sum_{j} \frac{\partial L_{\rm eff}}{\partial \dot{\ff m}_{j}} \ddot{\ff m}_{j}
  =
  \frac{d}{dt} \left(
  \sum_{j}
  \frac{\partial L_{\rm eff}}{\partial \dot{{\ff m}}_{j}} 
  \dot{{\ff m}}_{j}
  \right)
  \nonumber \\
  &=&
  \frac{d}{dt} \left(
  \sum_{j} \ff A(\ff m_{j}) S_{j} \dot{\ff m}_{j}
  +
  \sum_{i} \ff A(\ff n_{0}(\ff m)) s_{i} 
  \sum_{j} (\dot{\ff m}_{j} \nab_{j}) \ff n_{0,i}(\ff m)
  \right)
\ea
Inserting this in Eq.\ (\ref{eq:econs}) yields $dE/dt=0$.

Alternatively, total-energy conservation can be verified by using the adiabatic equation of motion (\ref{eq:asd}) and the sum rule (\ref{eq:sumrule}):
\ba 
  \frac{d}{dt} H_{\rm eff}(\ff m) 
  &=& 
  \sum_{j}
  \frac{\partial H_{\rm eff}}{\partial {\ff m}_{j}} 
  \dot{\ff m}_{j}
  =
  \sum_{j}
  \frac{\partial H_{\rm eff}}{\partial {\ff m}_{j}} 
  \frac{1}{S_{j}}  
  \left(
  \frac{\partial H_{\rm eff}}{\partial {\ff m}_{j}} \times \ff m_{j}
  +
  \ff T_{j} \times \ff m_{j}
  \right)
  =
  \sum_{j}
  \frac{1}{S_{j}}  
  \frac{\partial H_{\rm eff}}{\partial {\ff m}_{j}} 
  \cdot
  \ff T_{j} \times \ff m_{j}
  \nonumber \\
  &=&
  - \sum_{j}
  \frac{1}{S_{j}}  
  \frac{\partial H_{\rm eff}}{\partial {\ff m}_{j}} 
  \times 
  \ff m_{j}
  \cdot
  \ff T_{j} 
  =
  - \sum_{j}
  \left(
  \dot{\ff m}_{j}
  -
  \frac{1}{S_{j}}  
  \ff T_{j} \times \ff m_{j}
  \right)
  \cdot
  \ff T_{j} 
  = 0
  \: . 
\ea

\end{widetext}

\end{document}